\documentclass[aps, 10pt, prd, notitlepage,
twocolumn, superscriptaddress, nofootinbib, tightenlines]{revtex4-2}

\usepackage[linktocpage,breaklinks]{hyperref}
\usepackage[usenames,dvipsnames]{xcolor}
\usepackage{newtxtext}
\usepackage{newtxmath}
\usepackage[T1]{fontenc}
\usepackage{amsmath}
\usepackage{tensor}
\usepackage[utf8]{inputenc}
\usepackage{mathrsfs}
\usepackage{bm}

\usepackage{graphicx}
\usepackage{epsfig}
\usepackage{epstopdf}
\usepackage{natbib}
\usepackage[capitalise]{cleveref}
\usepackage{multirow}

\definecolor{romared}{RGB}{142,0,28}
\hypersetup{colorlinks=true,
            citecolor=romared,
            linkcolor=romared,
            urlcolor=romared}

\newcommand{\be}{\begin{equation}}
\newcommand{\ee}{\end{equation}}

\begin{document}
\title{Inspirals into bosonic dark matter stars and chirp mimickers}

\begin{abstract}
We investigate extreme--mass--ratio inspirals in which a stellar-mass compact object orbits a supermassive bosonic dark matter star, modeled as a boson star, using fully relativistic perturbative methods. 
Unlike inspirals around electro-vacuum black holes, these systems can shed scalar matter through dynamical friction which significantly alters the inspiral dynamics. 
We show that this additional dissipation can induce a chirp-like gravitational-wave signal closely resembling that of black hole binaries, allowing boson stars to act as gravitational-wave chirp mimickers even when they are not ultracompact. 
The inspiral evolution and resulting waveform depend sensitively on the compactness of the central boson star: highly compact configurations trigger dipolar scalar radiation, leading to a rapid plunge, whereas less compact stars yield smoother inspirals dominated by gravitational and quadrupolar scalar waves. 
To support waveform modeling, we derive semi-analytical prescriptions for the gravitational and scalar energy fluxes that remain accurate deep into the relativistic regime. 
Our findings indicate that future space-based detectors such as LISA could distinguish these mimicker signals from true black hole inspirals through measurable phase dephasings induced by scalar dissipation.

\end{abstract}

\author{Caio F. B. Macedo}
\email{caiomacedo@ufpa.br}
\affiliation{Faculdade de Física, Campus de Salin\'opolis, Universidade Federal do Par\'a,
Salin\'opolis, Par\'a, 68721-000 Brazil}

\author{Haroldo C. D. Lima}
\affiliation{Programa de Pós-Graduação em Física, Universidade Federal do Maranhão, Campus Universitário do Bacanga, 65080-805, São Luís, Maranhão, Brazil}

\author{ Raissa F. P. Mendes}
\affiliation{Instituto de Física, Universidade Federal Fluminense, Niterói, RJ, 24210-346, Brazil}
\affiliation{CBPF - Centro Brasileiro de Pesquisas Físicas, 22290-180, Rio de Janeiro, RJ, Brazil}

\author{Rodrigo Vicente}
\affiliation{Gravitation Astroparticle Physics Amsterdam (GRAPPA), University of Amsterdam,
Science Park 904, 1098 XH, Amsterdam, The Netherlands}

\author{Vitor Cardoso}
\affiliation{Center of Gravity, Niels Bohr Institute, Blegdamsvej 17, 2100 Copenhagen, Denmark}
\affiliation{CENTRA, Departamento de Física, Instituto Superior Técnico – IST,
Universidade de Lisboa – UL, Avenida Rovisco Pais 1, 1049 Lisboa, Portugal}

\date{{\today}}
\maketitle

%%%%%%%%%%%%%%%%%%%%%%%
\section{Introduction}
\label{sec:int}
%%%%%%%%%%%%%%%%%%%%%%%

\begin{figure*}
\includegraphics[width=0.8\linewidth]{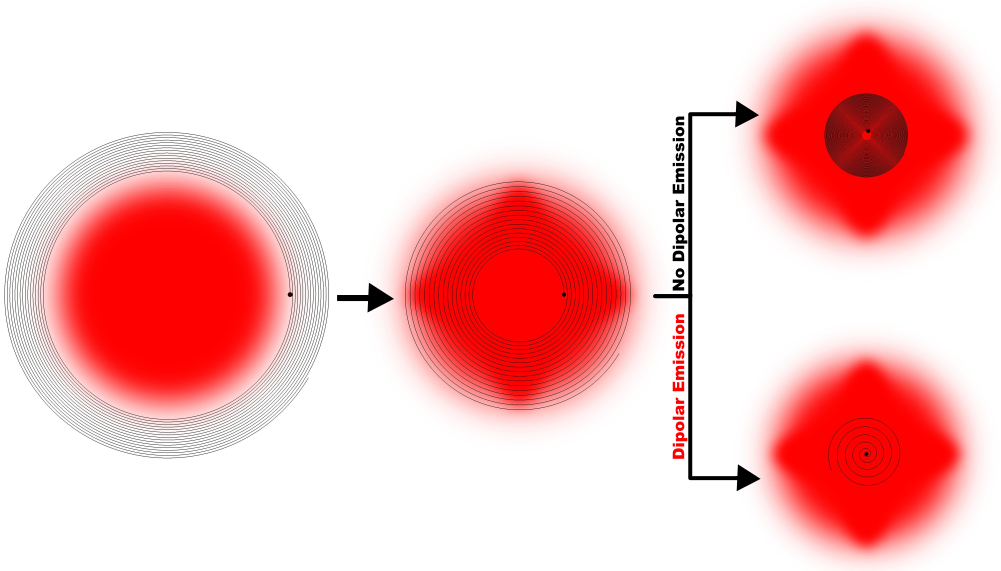}
    \caption{Schematic illustration of the evolution of an EMRI around a supermassive BS. In the initial stage (left panel), the evolution proceeds similarly to the black hole case, with the inspiral primarily driven by the emission of quadrupolar gravitational waves. As the smaller object enters the BS (middle panel), two effects take place: a resonant interaction that deposits energy into the star’s matter—altering its configuration—and the emission of scalar waves, often referred to as dynamical friction. These processes accelerate the inspiral, causing the smaller object to plunge more rapidly. Finally, depending on the compactness of the BS, two scenarios can occur. If the star is not compact enough for dipolar emission to arise (upper-right panel), the inspiral proceeds slowly toward the center, producing an almost monochromatic signal that depends on the star’s structure. Conversely, if the star is sufficiently compact (bottom-right panel), dipolar emission triggers a rapid plunge of the smaller compact object, abruptly quenching the radiation.}\label{fig:schematics}
\end{figure*}
The nature of dark matter (DM) remains one of the most enduring mysteries in cosmology and fundamental physics. A wide range of observations, from galactic rotation curves and galaxy cluster dynamics to gravitational lensing and the anisotropies of the Cosmic Microwave Background, indicate that nearly 85\% of the matter content of the Universe is composed of a non-luminous, non-baryonic component that clusters gravitationally~\cite{Zwicky1933,Rubin1970,Planck2018,Anderson2014,Ferreira2021}. Despite its success in describing the large-scale structure within the $\Lambda$CDM framework, the microscopic nature of DM remains unknown. Among the many candidates proposed, ultra-light bosonic fields---with masses in the range $10^{-24}\,{\rm eV} \lesssim \mu \lesssim 1\,{\rm eV}$---stand out as particularly well-motivated alternatives~\cite{Ferreira2021,Hui2017,Marsh2016}. For sufficiently small masses, such fields can condense into a single, coherent macroscopic state, forming a Bose--Einstein condensate or even a superfluid on galactic scales. The collective wave nature of such ultra-light dark matter models suppresses structure formation below a de~Broglie wavelength, thereby alleviating several long-standing small-scale tensions of the cold dark matter paradigm, such as the cusp--core and missing-satellite problems~\cite{Ferreira2021,Hu2000,Weinberg:2013aya,Schive2014}. Depending on the field self-interactions and initial conditions, these models manifest as \emph{fuzzy}, \emph{self-interacting fuzzy}, or \emph{superfluid} dark matter, each class predicting distinctive galactic-scale phenomenology~\cite{Ferreira2021}.

An intriguing consequence of ultra-light dark matter models is the possible formation of compact self-gravitating configurations known as \emph{boson stars} (BSs)~\cite{Kaup1968,Ruffini1969,Liebling2023,Schunck2003}. In the simplest scenario, BSs are stationary solutions of the Einstein--Klein--Gordon equations describing a complex scalar field bound by its own gravity. The competition between the gravitational attraction and the dispersive nature of the scalar field allows for stable equilibrium configurations, which can be interpreted as macroscopic quantum states. The maximum mass of such objects scales inversely with the boson mass, $M_{\rm max}\sim M_{\rm Pl}^2/m$~\cite{Liebling2023}, so that for $\mu\sim 10^{-10}\,{\rm eV}$ one obtains $M_{\rm max}\sim M_\odot$, while for $\mu\sim 10^{-22}\,{\rm eV}$ the masses can reach supermassive scales, $M_{\rm max}\sim 10^{9}M_\odot$. These configurations have therefore been proposed both as galactic dark matter cores and as supermassive compact alternatives to black holes at the centers of galaxies~\cite{Liebling2023,Schunck2003,Herdeiro2017}. 
Such scenarios open a new window for testing scalar-field dark matter through gravitational and electromagnetic observations, especially in the era of precision experiments targeting the strong-gravity regime.

There exist different ``flavors'' of BSs, depending on the self-interaction potential of the scalar field~\cite{Palenzuela:2006wp,Cardoso:2021ehg,Colpi:1986ye,Amaro-Seoane:2010pks}. These configurations can share many properties with black holes and have, in fact, been proposed as black-hole mimickers, particularly because they can reach supermassive scales. For instance, a BS located at the center of a galaxy has been shown to reproduce the same emission spectrum of accretion disks in certain models~\cite{Guzman:2005bs} and a similar silhouette (shadow-like) feature in compact cases~\cite{Rosa:2023qcv,Vincent:2015xta}. From the gravitational-wave perspective, these supermassive BSs can form extreme mass-ratio inspirals (EMRIs), exhibit resonant features that can induce detectable dephasings in the waveform, potentially observable by future space-based detectors such as LISA, TianQin, and Taiji~\cite{Macedo:2013jja,Macedo:2013qea}. Even beyond these resonant effects, the orbital motion within the star---which is allowed due to the minimal coupling between the scalar field and the orbiting matter---can give rise to distinctive gravitational-wave signatures~\cite{Kesden:2004qx}. It is worth noting, however, that additional dissipative mechanisms, such as accretion and dynamical friction, may play a crucial role in shaping the orbital evolution and the resulting gravitational-wave emission for trajectories within the star. These effects were not properly accounted for in previous works, considering compact BSs as the central object for EMRIs.

Dynamical friction can significantly modify the inspiral phase of an EMRI. A closely related scenario to the boson-star case is that of a supermassive black hole surrounded by bosonic matter fields, such as scalar clouds, as explored in relativistic scenarios in Refs.~\cite{Brito:2023pyl,Duque:2023seg,Li:2025ffh,Dyson:2025dlj}. In such systems, even though the ambient density is typically low, matter-induced effects can dominate over gravitational-wave dissipation in certain regimes. Within the LISA frequency band, for instance, where the evolution is generally driven by gravitational-wave emission, the influence of dynamical friction can still lead to detectable imprints of the surrounding environment~\cite{Duque:2023seg}. Environmental interactions can also strongly affect the eccentricity evolution of binaries and can impart a center-of-mass (CM) recoil similar to gravitational-wave kicks~\cite{Cardoso:2020lxx,Cardoso:2020iji}. In sufficiently dense media, these environmental effects may become strong enough to drive the inspiral rapidly, producing waveforms that resemble those from vacuum black-hole mergers~\cite{Roy:2024rhe}.

Although several studies have investigated supermassive BSs as black-hole mimickers, a detailed analysis of the inspiral evolution and the corresponding waveform remains largely unexplored.

Here we outline the main effects expected to drive an EMRI in which the central object is a BS. In particular, we evolve the binary under a quasi-circular approximation, accounting for both gravitational-wave and scalar-wave emission, the latter being typically associated with dynamical friction. The scalar energy fluxes are akin to the ionization studied in the Newtonian evolution of binaries around gravitational atoms~\cite{Baumann:2021fkf}. We show that, depending on the compactness of the star, the orbital evolution and the corresponding waveform can differ significantly, mainly due to the possible absence of dipolar emission in certain configurations. In some cases, a chirp-like feature may arise as a result of a rapid plunge driven by intense dipolar scalar emission. This suggests that ordinary BSs could act as black hole mimickers in binary evolution, even when they are not ultracompact. A summary of the possible evolutions of EMRI into BSs can be seen in Fig.~\ref{fig:schematics}.

The remainder of this work is organized as follows. In Sec.~\ref{sec:setup}, we describe the main setup, namely the Einstein–Klein–Gordon system with a perturbing particle. We also present the computation of circular geodesics used to track the particle’s motion. In Sec.~\ref{Sec:Flux}, we compute the energy fluxes emitted by a system consisting of a BS orbited by a massive particle, and we introduce a modified Newtonian approach that simplifies the full numerical computation. We further describe the procedure for evolving the motion in the quasi-circular regime and for computing the waveform at leading (quadrupolar) order. In Sec.~\ref{sec:numericalresults}, we present our numerical results, including the inspiral evolution and representative examples of waveforms. Finally, in Sec.~\ref{sec:conclusion}, we summarize our main conclusions. Throughout this work, we adopt natural units with $G=c=1$.

%%%%%%%%%%%%%%%%%%%%%%%%%%
\section{General setup}\label{sec:setup}
%%%%%%%%%%%%%%%%%%%%%%%%%%

\subsection{Background and perturbations}

We consider a setup consisting of a $U(1)$-invariant complex scalar field minimally coupled to gravity. The dynamics of the system is governed by the following action
\begin{align}
	S\!=\!\int \! \left[\frac{R}{16\pi} - \frac{1}{2} g^{\mu\nu} \nabla_\mu\Phi^* \nabla_\nu\Phi - \frac{1}{2} V(|\Phi|^2)\right] \sqrt{-g}\,d^4x+S_m,
\label{eq:ActionScalar}
\end{align}
where $R$ is the Ricci scalar associated with the metric $g_{\mu\nu}$, $g$ is the metric determinant, $\Phi$ is a $U(1)$-invariant and minimally coupled complex scalar field, and $V(|\Phi|^2)$ denotes its self-interaction potential. The matter sector is described by the action $S_m$. Our analysis does not strongly rely on the specific form of the potential $V(|\Phi|^2)$. However, for the numerical results presented in this work, we adopt the quadratic potential $V(|\Phi|^2)=\mu^2|\Phi|^2$, where $\mu$ is associated with the mass of the scalar field. 

Varying the action with respect to the metric and the scalar field yields the field equations
\begin{align}
G_{\mu\nu}&=8\pi T_{\mu\nu},\label{eq:einstein}\\
~\nabla_\mu\nabla^\mu\Phi&=\frac{dV(|\Phi|^2)}{d|\Phi|^2}\Phi,\label{eq:scalar}
\end{align}
with $G_{\mu\nu}$ denoting the Einstein tensor and 
\begin{equation}
    T_{\mu\nu} = T_{\mu\nu}^\Phi + T_{\mu\nu}^m,
\end{equation}
where $T_{\mu\nu}^m$ is the energy-momentum tensor derived from $S^m$ and 
\begin{align} \label{eq:TmunuPhi}
 T_{\mu\nu}^\Phi= \frac{1}{2} \left[\nabla_\mu\Phi \nabla_\nu \Phi^* + \nabla_\nu\Phi \nabla_\mu \Phi^* - g_{\mu\nu}\left( \nabla^\alpha\Phi^*\nabla_\alpha\Phi+V \right) \right]
\end{align}
is the energy-momentum tensor associated with the scalar field.

Owing to the $U(1)$-invariance of the complex scalar field and Noether's theorem, the theory admits a conserved current, 
\begin{align}
j_\mu=i\left(\Phi\nabla_\mu\Phi^*-\Phi^*\nabla_\mu\Phi \right),    
\end{align}
which satisfies the continuity equation, $\nabla_\mu j^\mu=0$. This allows us to define the conserved scalar quantity
\begin{align}
Q=\int_{\Sigma_t} j^\mu n_\mu d\Sigma,    
\end{align}
where $\Sigma_t$ is a spacelike hypersurface with future-pointing normal vector $n_\mu$. The conserved quantity $Q$ represents the total number of scalar particles in the system.

For the background spacetime, we consider that no matter fields other than the scalar field are present, and assume a static, spherically symmetric configuration described by
\begin{align}
\label{background} ds^2=-A(r)\,dt^2+\frac{dr^2}{B(r)}+r^2d\Omega^2, \quad \Phi(t,r)=\phi_0(r)\,e^{i\omega t}, 
\end{align}
where $A(r)$, $B(r)$ and $\phi_0(r)$ are functions of the radial coordinate determined by the field equations and the boundary conditions, and $d\Omega^2$ is the line element of the unit 2-sphere. Although the scalar field is assumed to oscillate with frequency $\omega>0$, its energy-momentum tensor (\ref{eq:TmunuPhi}) has no time dependence, consistent with the assumption of a static background. 
In Appendix~\ref{app:bosonstar}, we describe the numerical techniques employed to solve the field equations~\eqref{eq:einstein} and \eqref{eq:scalar}, and to compute the background functions in Eq.~\eqref{background}. 

We then introduce a point-particle perturber with rest mass $m_p$ on the background~\eqref{background}, so that\footnote{The 4-dimensional Dirac delta distribution is defined by $F(\bar{x})\equiv \int F(x)\delta^{(4)}(x-\bar{x})\,d^4x$ where $F$ is any scalar function of the spacetime coordinates. Since the invariant volume element in a (pseudo)-Riemannian manifold is $\sqrt{-g}\, d^4x$, the corresponding invariant Dirac delta distribution is expressed as $\delta^{(4)}(x-\bar{x})/\sqrt{-g}$. The stress-energy tensor of the point-particle~\eqref{eq:Tpp} is formulated using the invariant Dirac delta distribution.}
\begin{equation}
    \mathcal{L}^m = \int m_p\, \frac{\delta^{(4)}[x-x_p(\tau) ]}{\sqrt{-g}}\, d\tau,
\end{equation}
with $x^\mu_p(\tau)$ denoting the particle worldline, parametrized by proper time $\tau$.
The particle mass is assumed to be much smaller than that of the central BS, $M$, placing the system within the regime of an EMRI. The stress-energy tensor of the point particle takes the standard form
\begin{align}
\label{eq:Tpp} T^{\mu\nu}_p=\int m_p\,\dot{x}_p^\mu(\tau)\,\dot{x}_p^\nu(\tau) \frac{\delta^{(4)}[x-x_p(\tau) ]}{\sqrt{-g}}\, d\tau ,
\end{align}
where $\dot{x}_p^\mu=dx_p^\mu/d\tau$ denotes the particle's four-velocity. In the test-particle limit, the perturber follows geodesic motion. The presence of the point particle perturbs the background configuration according to
\begin{align}
&g_{\mu\nu} \rightarrow g_{\mu\nu}+q\,\delta g_{\mu\nu}+\mathcal{O}(q^2), \\
& \Phi \rightarrow \Phi+ q\,\delta \Phi+\mathcal{O}(q^2),
\end{align}
where $q\equiv m_p/M$ is the ratio between the point particle mass and that of the BS. We show the main equations to be integrated in Appendix~\ref{app:bosonstar} and~\ref{app:perturbations}, but also refer the reader to Refs.~\cite{Macedo:2013jja,Duque:2023seg} for more details.

\subsection{Geodesics around a boson star}
\begin{figure}
    \centering
    \includegraphics[width=1\linewidth]{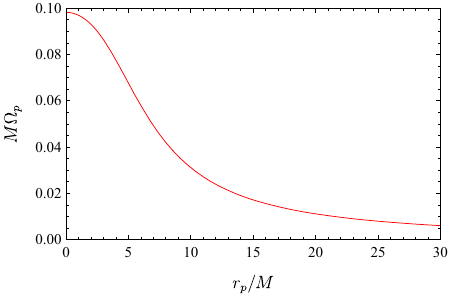}
    \caption{Orbital frequency of the perturber on a circular geodesic as a function of the radial coordinate $r_p$. The curve corresponds to the most compact BS configuration, with total mass $M\mu=0.633$.}
    \label{fig:orbital_fre}
\end{figure}

To leading order in $q$, the perturber approximately follows a timelike geodesic in the background spacetime of the BS. These geodesics can be analyzed using the Lagrangian formalism, with the Lagrangian given by: 
\begin{align}
\label{eq: Hamiltonian}{\tt L} \equiv \frac{m_p}{2}g_{\mu\nu}\dot{x}^\mu\,\dot{x}^\nu,
\end{align}
with $\dot{x}^\mu = dx^\mu(\lambda)/d\lambda$. Without loss of generality, we can restrict our attention to the $\theta=\pi/2$ plane, due to the invariance of the Lagrangian under the action of the $SO(3)$ group, inherited from the spherical symmetry of the background. 
The Euler-Lagrange equations,
\begin{align}
\frac{\partial {\tt L}}{\partial x^\mu}-\frac{d}{d\lambda}\left(\frac{\partial {\tt L}}{\partial \dot{x}^\mu} \right)=0, 
\end{align}
for the coordinates $t$ and $\varphi$, lead to conserved quantities
\begin{align}
\label{eq:dotx}g_{tt}\,\dot{t}=\frac{E}{m_p}, \quad g_{\varphi\varphi}\,\dot{\varphi}=\frac{L}{m_p}, 
\end{align}
where $E$ and $L$ are the energy and angular momentum of the perturber, as measured by an observer at spatial infinity. Substituting Eq.~\eqref{eq:dotx} into Eq.~\eqref{eq: Hamiltonian}, and using the normalization of the four-velocity ($g_{\mu\nu}\dot{x}^\mu\dot{x}^\nu=-1$), one finds that
\begin{align}
AB^{-1}\,m_p^2\,\dot{r}^2+V(r)=E^2,    
\end{align}
where $V(r)$ represents an effective potential for timelike geodesics, given by
\begin{align}
V(r)= A\left(\frac{L^2}{r^2}+m_p^2 \right).  
\end{align}
We shall focus on scenarios in which the inspiral proceeds through a sequence of quasi-circular orbits. Circular orbits of radius $r_p$ are defined by the conditions $\dot{r}=0$ and $\ddot{r}=0$, which are equivalent to
\begin{align}
\label{eq:circ_orbit}V(r_p)=E^2, \qquad V'(r_p)=0,   
\end{align}
with a prime denoting a derivative with respect to $r$.
From Eq.~\eqref{eq:circ_orbit}, we can obtain the energy and angular momentum of the perturber at a radius $r_p$:
\begin{align}
\label{eq:energy_L}
&\frac{E^2_{\text{orbit}}}{m_p^2}=\frac{2\,A(r_p)^2}{2 A(r_p)-r_p\,A'(r_p)},\\
&\frac{L^2_{\text{orbit}}}{m_p^2}=\frac{r_p^3A'(r_p)}{2A(r_p)-r_p\,A'(r_p)}.
\end{align}
Moreover, the orbital angular frequency of the perturber can be determined through
\begin{align}
\Omega_p=\left. \frac{d\varphi}{dt} \right|_{r_p}= 
\sqrt{\frac{A'(r_p)}{2\,r_p}}.    
\end{align}
Figure~\ref{fig:orbital_fre} shows the orbital angular frequency as a function of the orbital radius $r_p$ for a BS with mass $M\mu=0.633$. This curve is representative of all BSs on the stable branch, with the maximum frequency $\Omega_{p, \rm max}$ occurring at $r_p\approx 0$. As we shall see, the value of $\Omega_{p, \rm max}$ can be used as a discriminator for the evolution of the inspiral into a BS.

%%%%%%%%%%%%%%%%%%%%%%%%%%%%%%%%%%
\section{Flux computation and the semi-analytical approach}
\label{Sec:Flux}
%%%%%%%%%%%%%%%%%%%%%%%%%%%%%%%%%
%
\begin{figure}
    \centering
    \includegraphics[width=1\linewidth]{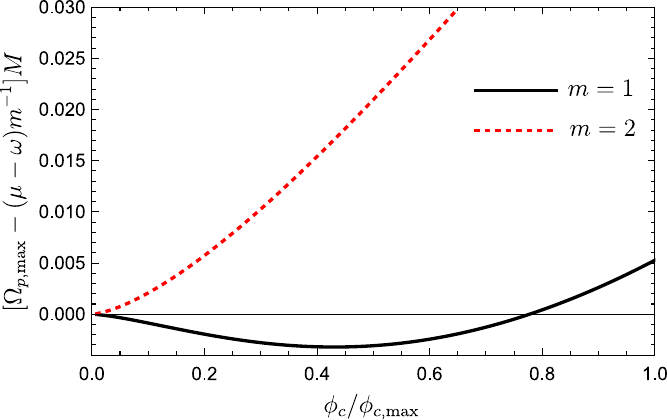}
    \caption{Threshold for dipolar and quadrupolar scalar emission. The (scaled) left-hand side of Eq.~\eqref{ineq:scalar_rad} is shown as a function of the central value of the scalar field. When this quantity is positive, scalar emission is allowed for a given multipole.
    Note that while quadrupolar emission is possible for any BS configuration, dipolar emission only occurs for the largest values of the central scalar field, corresponding to highly compact solutions.}
    \label{fig:condition}
\end{figure}
\begin{figure}
    \centering    \includegraphics[width=1\linewidth]{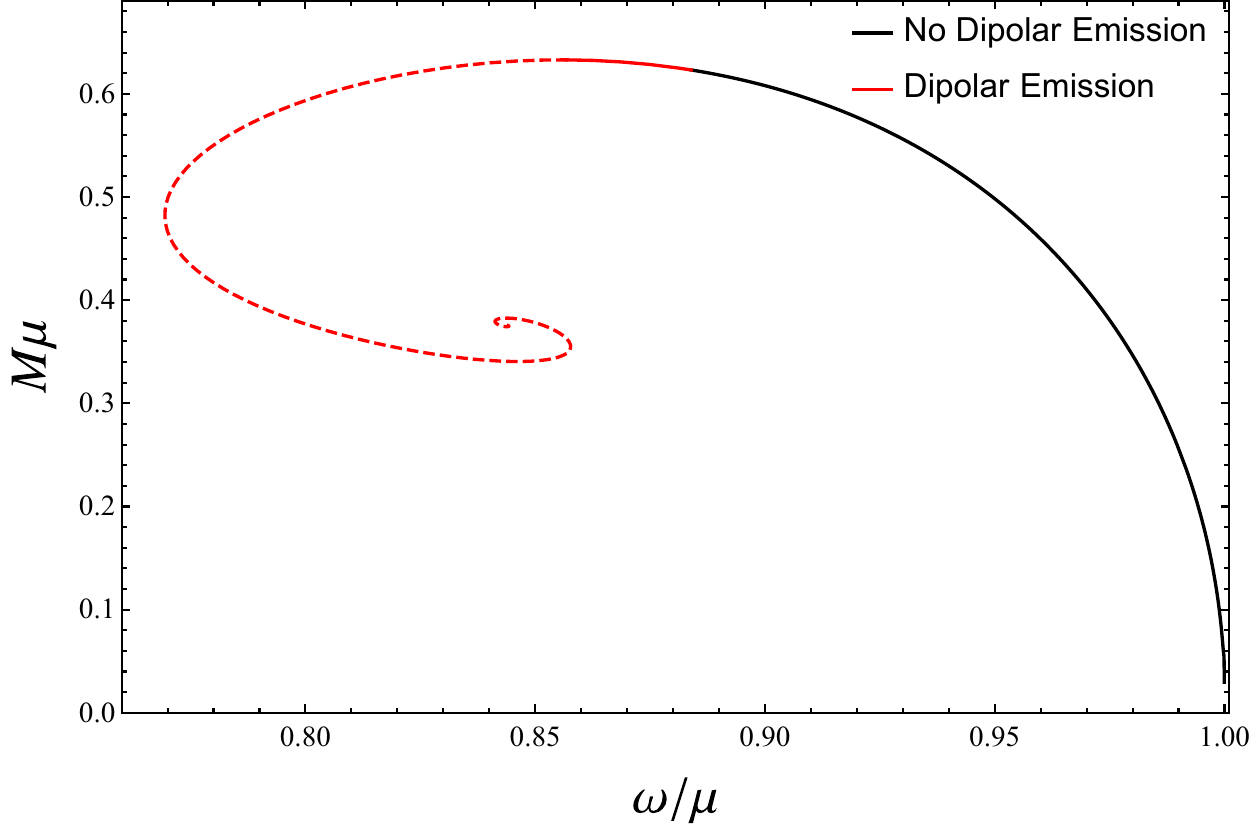}
    \caption{The total mass of the BS as a function of the scalar field frequency. Dipolar emission by an orbiting point particle is absent in the black branch and can be present in the red branch. The dashed branch denotes unstable BS solutions.}
    \label{fig:Mw}
\end{figure}
The introduction of a point-particle perturber into the background of a BS triggers the emission of both gravitational and scalar waves. As a result, the system loses energy and the particle inspirals toward the center of the BS. At leading order in the perturbation, considering the EMRI regime, the energy emitted over a single orbit is much smaller than the total energy of the particle. Therefore, we can assume that the perturber inspirals adiabatically, i.e., through a sequence of quasi-circular orbits with progressively decreasing radius.

To describe the inspiral, we need to compute the gravitational, scalar, and particle number fluxes. For this purpose, we first determine the linear perturbations of the metric and scalar field induced by a particle in circular orbit, and then evaluate the fluxes through~\cite{Brito:2023pyl,Duque:2023seg,Duque:2024fot} (see also Appendix~\ref{app:perturbations} for details) 
\begin{align}
\label{eq:flux_Eg}  ^{\ell m} \dot{E}^g_{\infty}=&\frac{1}{64\pi}\frac{(\ell+2)!}{(\ell-2)!}m^2\Omega_p^2|^{\ell m}\psi_{Z,\infty}|^2,~~\ell+m~~{\rm even},\\
\label{eq:flux_EPhi}  ^{\ell m}\dot{E}^\Phi_{\infty}=&(\omega+m\Omega_p){\rm Re}\left[\sqrt{(\omega+m\Omega_p)^2-\mu^2}\right]|^{\ell m}\phi_{+,\infty}|^2,\\
\label{eq:flux_QPhi}  ^{\ell m}\dot{Q}^\Phi_{\infty}=&{\rm Re}\left[\sqrt{(\omega+m\Omega_p)^2-\mu^2}\right]|^{\ell m}\phi_{+,\infty}|^2,\\
\label{eq:flux_EBS}  ^{\ell m} \dot{E}^{\rm BS}_{\infty}=&-(\omega+\omega' Q^{\rm BS})\, ^{\ell m}\dot{Q}^\Phi_{\infty},
\end{align}
where $^{\ell m} \dot{E}^g_{\infty}$ is the gravitational energy flux, $^{\ell m}\dot{E}^\Phi_{\infty}$ the scalar field energy flux, $^{\ell m}\dot{Q}^\Phi_{\infty}$ the particle number flux, and $^{\ell m} \dot{E}^{\rm BS}_{\infty}$ the rate of change in the BS energy, {with $\omega'\equiv d \omega/d Q^{\rm BS}$. The term proportional to $\omega'$ in \eqref{eq:flux_EBS} originates from the secular evolution of the BS with changing $Q^{\rm BS}$ (see Appendix~\ref{app:backreaction}).} Each of these contributions is due to a $(\ell,m)$ multipole. The quantities $^{\ell m}\psi_{Z,\infty}$ and $^{\ell m}\phi_{+,\infty}$ are the gravitational (Zerilli) and scalar perturbation amplitudes computed at infinity, respectively, obtained by solving the perturbation equations, which we describe in Appendix~\ref{app:perturbations}. We shall focus on the dominant multipoles, i.e. $\ell=m$. We also neglect the axial sector, which contributes subdominantly to the gravitational wave emission and does not excite scalar radiation~\cite{Macedo:2013jja}. The total fluxes---i.e. the summation of all multipoles---will be represented by the same symbols as above, but without the index $\ell m$, e.g. $E^g_\infty=\sum_{\ell m}(^{\ell m}E_{\infty}^g)$.

It is important to note that scalar emission is not allowed at all frequencies. The condition for scalar emission to occur follows from requiring a wave-like behavior at infinity, which (after simplifications) reduces to
\begin{align}
\label{ineq:scalar_rad} \omega+m\,\Omega_p-\mu >0. 
\end{align}
For a given BS solution with frequency $\omega$, we can compute the maximum orbital frequency $\Omega_{p, \rm max}$ (typically attained at $r_p \sim 0$) and verify whether or not the inequality \eqref{ineq:scalar_rad} is satisfied. This result defines the threshold for scalar emission for a given multipole.

In Fig.~\ref{fig:condition} we show the left-hand side of Eq.~\eqref{ineq:scalar_rad} for the maximum orbital frequency $\Omega_{p, \rm max}$, and two multipoles, $m=1$ and $m=2$, as a function of the central scalar field of the BS. We see that dipolar emission occurs only in configurations with high central field values, which correspond to highly compact BSs. On the other hand, quadrupolar (and also $\ell\geq3$) emission is always possible, regardless of the BS compactness. To emphasize this point, in Fig.~\ref{fig:Mw} we present the $(\omega/\mu,M\mu)$ diagram for BSs, choosing the color scheme to highlight the branches where dipolar emission is forbidden (black branch) or allowed (red branch). Dipolar emission sets in for higher-mass BSs, with the threshold located at $(\omega/\mu \sim 0.88,\  M\mu\sim 0.62 )$. We observe that this configuration is located at the stable branch. Therefore, there exists a subset of stable BS solutions that allows for dipolar emission, represented by the solid red branch in Fig.~\ref{fig:Mw}. The dashed red branch corresponds to unstable BSs capable of emitting dipolar radiation. 

%%%%%%%%%%%%%%%%%%%%%%%%%%%%%%%%%%%%%%%%%%%%%%
\subsection{A Newtonian prescription for the fluxes}
%%%%%%%%%%%%%%%%%%%%%%%%%%%%%%%%%%%%%%%%%%%%%%%
We can also, for simplicity, employ a semi-analytic approximation for the gravitational flux by considering the weak-field (Newtonian) formula: 
\begin{equation} \label{eq:NGW_flux}
    \dot{E}_{\rm N}^g= \frac{32}{5}m_{p}^2 r_{p}^4 \Omega_{p}^6+\frac{62}{7}m_{p}^2 r_{p}^6\Omega_{p}^8,
\end{equation}
where the first term on the right-hand side corresponds to the quadrupole contribution, and the second to the octupole. While this formula provides a good approximation for the early inspiral phase, it fails to accurately capture the full dynamics in the regime $r_p \sim \mathcal{O}(R)$, which is the main focus of this work. In this regime, we find that a good eyeballing adjustment is provided by the heuristic expression
\begin{align}
\label{eq:mod_flux} \dot{E}^{g}_{\infty}\approx|g_{tt}(r_p)|^\alpha \dot{E}_{\rm N}^g,
\end{align}
in which $\alpha \sim  1$ is found to be a good adjustment for all stable solutions.  The term $|g_{tt}(r_p)|^\alpha$ in Eq.~\eqref{eq:mod_flux} accounts for the fact that the Newtonian expression (\ref{eq:NGW_flux}) for the gravitational flux is larger than the result obtained by numerically solving the perturbation equations.

To compute the scalar flux, we need to obtain the field perturbation $\delta \Phi$ at infinity (cf. Appendix~\ref{app:perturbations}). This requires solving the full set of coupled equations, which can be done, for instance, using the method of variation of parameters. However, a semi-analytic expression can be obtained by considering the Newtonian approximation~\cite{Annulli:2020lyc,Duque:2023seg}
\begin{equation}
    \label{newt_phi}^{m}\phi_{+,\rm N}= - \frac{8\pi^{\frac{3}{2}} m_p \phi_0(r_p) m^{\frac{m}{2}-1} Y_m^m\left(\frac{\pi}{2},0\right) (r_p \Omega_p)^m}{2^{\frac{m}{2}+2} (\Omega_p/\mu)^{\frac{m}{2}+1} \Gamma\left(m+\frac{3}{2}\right)},
\end{equation}
which is valid for $m\geq2$, and where we have considered the dominant multipole, $\ell=m$.
Using the above expression, we can compute the scalar flux through Eq.~\eqref{eq:flux_EPhi}, using $^m\phi_+\approx {}^m\phi_{+,\rm N}$.
Although this provides a good approximation for Newtonian BSs, it fails to reproduce some features of higher compactness configurations, which are the main target of our analysis. As with the gravitational-wave flux, we find that a good adjustment to better match the strong field results is given by
\begin{equation}
\label{eq:scalar_flux}^{m}\dot{E}^\Phi_{\infty}=\frac{^{m}\dot{E}_{\rm N\Phi}}{|g_{tt}(r_p)|^{\beta}},
\end{equation}
where $\beta \sim 3/2$ yields good agreement with the numerical data, and $^{m}\dot{E}_{\rm N\Phi}$ gives the flux computed with the Newtonian expression for the scalar perturbation, Eq.~\eqref{newt_phi}. {The particle number flux also scales in the same way.}

With the improved analytical approximations for the gravitational, scalar and particle number fluxes, we can extend the analysis of BS emission properties to the fully relativistic regime, once the background is computed. In particular, Eq.~\eqref{eq:scalar_flux} enables us to investigate whether dipolar ($\ell=m=1$) and quadrupolar ($\ell=m=2$) scalar emissions are possible within the family of BS solutions.
While we shall not rely on the analytical expressions above to compute the evolution of the binaries, they provide a convenient tool to assess relevant features in future works. To illustrate the validity of the Newtonian approximations, we show a comparison between the numerical results and the analytical formulas in Appendix \ref{app:approximation}.

\subsection{Resonant configurations}

{
As the orbiting particle moves closer to the BS, some of the emitted radiation frequencies may approach the characteristic oscillation frequencies of the background configuration. In this regime, the system behaves effectively as a forced coupled oscillator, and the external driving source can trigger resonant responses~\cite{Pons:2001xs,Macedo:2013jja}. For a given azimuthal number $m$, the emitted wave frequency is $m\Omega_p$, so that resonances are expected whenever
\begin{equation}
    m \Omega_p+\omega \approx \omega_R,
\end{equation}
with $\omega_R$ being the real part of the relevant quasinormal-mode frequency of the boson star.\footnote{We notice that the resonant condition is shifted due to the background oscillation frequency of the BS. See Ref.~\cite{Macedo:2013jja} for details.} Consequently, as the particle crosses orbital radii in which the emitted frequency matches the resonant condition, the radiated fluxes are enhanced. The finite width of the resonance is set by the damping timescale of the mode, and is therefore controlled by the imaginary part of its frequency.
}

{BSs possess a sector of their spectrum that closely resembles that of the hydrogen atom. These modes are usually referred to as quasibound states, since the bosonic perturbations are exponentially damped at radial infinity. Previous works have shown that several of these modes can be excited during the evolution of inspirals in BS spacetimes (see, e.g.~\cite{Macedo:2013jja}). Moreover, the saturation of the threshold condition~\eqref{ineq:scalar_rad} appears to identify a special frequency around which many such excitations occur. We note that, because of the nature of these resonances, enhancements in the fluxes occur only in the gravitational radiation, through the coupling to the scalar sector, since the scalar perturbations are exponentially suppressed at infinity. Nevertheless, resonant configurations do induce a multipolar structure within the BS by enhancing the field perturbations in its interior.
}

\subsection{Inspiral evolution and waveform}
\label{Sec:Fluxes}

We now turn to analyze the orbital evolution of the point particle around the central BS. In the adiabatic regime, where the motion can be approximated as a sequence of quasi-circular geodesics (see Appendix~\ref{app:qc} for an assessment of the validity of this assumption), the orbital evolution is obtained through energy conservation:
\begin{align}\label{eq:flux}
\dot{E}_{\rm orbit}&=-{\cal F},~{\rm with}\\
{\cal F}&=\underbrace{\dot{E}^g_{\infty}}_{\rm GW}+\underbrace{\dot{E}^\Phi_{\infty}+ \dot{E}^{\rm BS}_{\infty}}_{\Phi},\nonumber
\end{align}
where $\dot{E}_{\rm orbit}$ is the rate of energy loss of the point particle. We highlight with the underbrace the contribution from the gravitational wave (GW) and the scalar field ($\Phi$) sectors. 
The orbital energy of the point particle in the BS spacetime is computed within the relativistic framework, as given by Eq.~\eqref{eq:energy_L}, under the assumption of circular orbits. %\cm{talk about the kludge approximation}
Therefore, in this approach, the particle is treated as evolving through a sequence of geodesics, with the orbital parameters slowly varying under the influence of the energy loss given by ${\cal F}$. {As the particle inspirals, its monopole field will additionally contract quasi-statically the BS, changing its binding energy in a way that cannot be directly quantified via asymptotic fluxes. We neglect that contribution here, as we expect it to be degenerate with the particle mass (see Appendix~\ref{app:backreaction} for an explicit computation in the Newtonian limit).}

\begin{figure*}
    \centering
    \includegraphics[width=0.8\columnwidth]{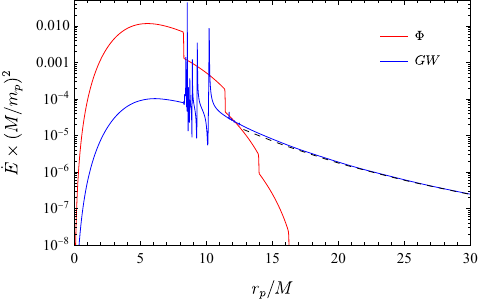}
    \includegraphics[width=0.8\columnwidth]{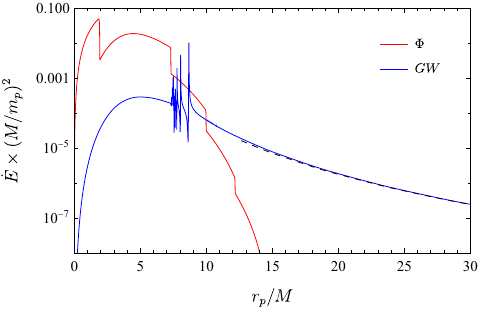}
    \caption{Scalar and gravitational-wave contributions to the rate of change of the particle's binding energy, as function of its radial position. We show the contribution of each part (GW and $\Phi$) as highlighted in Eq.~\eqref{eq:flux}. Left panel: Configuration with $M\mu=0.62448$, for which dipolar emission is absent. Right panel: the most massive configuration, with $M\mu=0.633$, is considered, which allows for dipolar emission for sufficiently low values of $r_p$. {The dashed line corresponds to the Newtonian expression.}}
    \label{fig:fluxE}
\end{figure*}

\begin{figure*}
    \includegraphics[width=0.8\columnwidth]{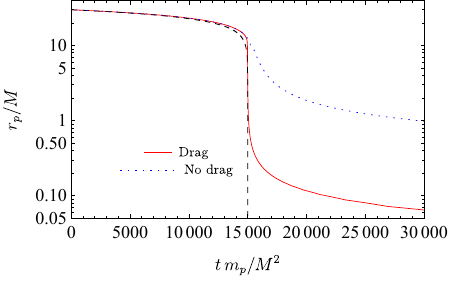}
    \includegraphics[width=0.8\columnwidth]{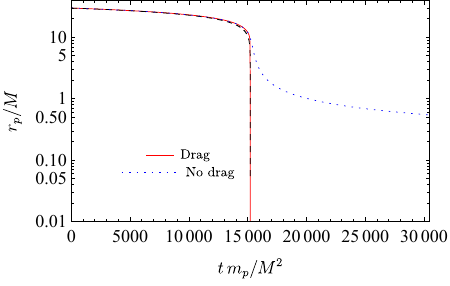}
     \caption{Orbital radius as function of time. We compare the inspiral evolution with (red curve) and without (blue dotted curve) the scalar radiation contribution to the energy flux. For reference, we include the result from the quadrupole approximation to a binary system of point particles, $\propto (1-t/t_c)^{1/4}$ (black dashed line), where $t_c$ is chosen to fit the numerical evolution. Left panel: Configuration with $M\mu=0.62448$, for which dipolar emission is absent. Right panel: Configuration with $M\mu=0.633$, corresponding to the most compact stable BS, and featuring dipolar scalar emission.}\label{fig:radius_evol}
\end{figure*}

We can now compute the time evolution of both the phase and orbital radius. The radial evolution follows from
\begin{equation}\label{eq:drorbit}
    \dot{r}_p=\dot{E}_{\rm orbit}\left(E_{\rm orbit}'\right)^{-1}=-\frac{\cal F}{E_{\rm orbit}'},
\end{equation}
with a prime denoting a radial derivative and a dot denoting a proper time derivative. To obtain Eq.~\eqref{eq:drorbit} we have used \eqref{eq:flux}. The phase $\Psi$ of the GW is given by~\cite{Yunes:2011aa} 
\begin{equation}
        \frac{d\Psi}{dr}= \frac{2\Omega_p}{\dot{E}_{\rm orbit}}E_{\rm orbit}'=-2\Omega_p\frac{E_{\rm orbit}'}{{\cal F}}.
\end{equation}
The radial derivative of the orbital energy can be obtained as function of the metric using Eq.~\eqref{eq:energy_L}. With these ingredients, we can construct the leading quadrupole Newtonian waveform. This is imported from the weak-field approximation, although we consider the orbital dynamics to be described by the relativistic model with all multipoles. This hybrid approach captures the essential features of the waveform (see also~\cite{Babak:2006uv,Levati:2025ybi}). The time-evolved amplitude is given by~\cite{Macedo:2013qea} 
\begin{align}
    h\approx\frac{4}{r}\frac{m_p}{M}\Omega_p^2[r_p(t)] r_p(t)^2\cos[\Psi(t)].
\end{align}
With the above, we are able to construct the GW waveforms for EMRIs around BSs within the perturbative framework. Far from the star, the formalism naturally reproduces the results of GR in a Schwarzschild spacetime, as the energy fluxes become very similar. As the orbiter gets close to the BS, resonant effects introduce additional dephasing~\cite{Macedo:2013qea,Macedo:2013jja}, and the scalar flux becomes dominant. This leads to a pronounced drag effect on the inspiral, significantly altering the orbital evolution. We will explore these effects in detail in the following section.

During the adiabatic inspiral, scalar particles—associated with the scalar field $\Phi$—are emitted due to the presence of the perturber. The total number of scalar particles radiated over the course of the inspiral, as a function of the radial distance, can be estimated as
\begin{equation}
\label{eq:Q_ejected}    \Delta Q^{\Phi}=\int_{t_i}^{t_f}\dot{Q}_\infty^{\Phi}dt=\int_{r_i}^{r_f}\frac{\dot{Q}_\infty^\Phi}{\dot{r}_p}dr=\int_{r_i}^{r_f}\frac{\dot{Q}_\infty^\Phi}{\dot{E}_{\rm orbit}}E'_{{\rm orbit}}dr,
\end{equation}
where $\dot{Q}_\infty^\Phi$ is the sum of the particle flux over the various multipoles. Tracking the total number of particles ejected is crucial, as this allows us to assess whether the BS changes considerably during the inspiral.

\section{Numerical results}\label{sec:numericalresults}

\subsection{Inspiral evolution and chirp mimickers}

We have numerically computed the evolution of a particle plunging into a BS using the procedures outlined in the previous section. To be pragmatic, we focus on the case where the particle starts on a circular orbit at $r_p/M\approx30$, which is distant enough so that the energy flux is basically the same as in Schwarzschild spacetime, and the early inspiral proceeds accordingly. In Fig.~\ref{fig:fluxE}, we present the fluxes for the two BS configurations analyzed in this work. The first corresponds to the marginally stable, maximum mass configuration, with $\mu M=0.633$, while the second has a comparable compactness, with $\mu M=0.62448$, but lies below the threshold for dipolar emission. The corresponding radius are $R=12.41M$ and $R=14.71M$, for $\mu M=0.633$ and for $\mu M=0.62448$, respectively, such that the BS are similar in compactness. 

{The taxonomy of the flux as a function of the orbital radius is similar in both cases. At large radii, the flux is dominated by gravitational-wave emission and is accurately described by the Newtonian approximation, shown by the dashed lines in Fig.~\ref{fig:fluxE}. As the particle approaches the star, the orbital frequency increases and progressively activates additional multipolar scalar channels, starting from higher values of $\ell$. This behavior is reflected in the step-like profile of the red curve in the figure. Whenever a scalar channel is near of being activated, a multitude of modes can be excited through a resonant process, leaving an imprint on the gravitational sector through the coupling between perturbations, which can be seen as the narrow peaks in Fig.~\ref{fig:fluxE}. This imprint appears only for $\ell\geq2$, with the most prominent contribution coming from the quadrupolar mode, $\ell=2$.
}

The results show that dipolar scalar flux introduces considerable dissipation within the star, corresponding to a stronger drag force on the orbiting particle. As we will see, this enhanced dissipation causes two otherwise similar BSs to generate noticeably different inspiral dynamics.

With the fluxes computed, we can proceed to numerically analyze how the orbital radius evolves during the inspiral. As an illustration of the main effects, we consider equal-mass binaries, but we also rescale the time parameter to reinforce that the overall qualitative results remain the same for other configurations.\footnote{We note here that in this picture the effect of choosing a different mass ratio would be a scaling factor of the plunging time in the radius evolution, which is accounted in the way we display our results. However, the waveforms shown will be different, as smaller mass-ratio will include many more cycles and a bigger dephasing between the cases, as we shall explain later on.} In Fig.~\ref{fig:radius_evol} we show the orbital radius as a function of time for the two BS fluxes shown in Fig.~\ref{fig:fluxE}. The quadrupolar approximation for the case of two point particles is shown as a dashed line for comparison. Both evolutions begin at $r_p/M = 30$ and initially follow a similar trend, with the orbital radius gradually decreasing.
In the left panel, corresponding to $\mu M=0.62448$, where dipolar radiation is absent, there is a sharp decrease in the orbital radius followed by a quasi-circular inspiral deep inside the star. In contrast, the right panel ($\mu M = 0.633$), which features dipolar emission, displays a more abrupt inspiral. Dipolar radiation enhances the drag force due to the scalar matter, accelerating the plunge and suppressing the final quasi-circular phase observed in the previous case. Still, in both cases, a steep radial plunge-like profile is present. 

\begin{figure*}
    \centering
    \includegraphics[width=.4\linewidth]{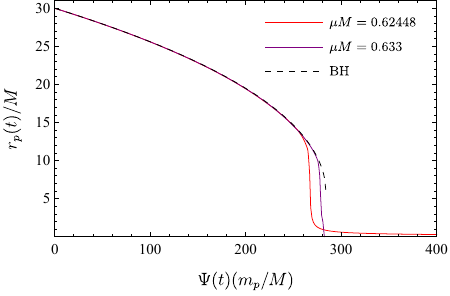}\includegraphics[width=.4\linewidth]{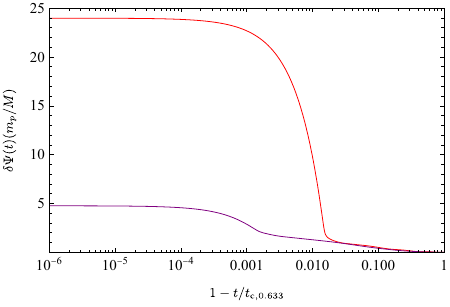}
    \caption{\textit{Left panel:} $(\Psi(t),r_p(t))$ plane as the inspiral evolves. We show the BS cases explored in this work and compare them to the BH case (up to $r_p=6M$), all scaled in the same way. They share a similar evolution, with the coalescence depending on the BS radius. \textit{Right panel:} {We show the phase difference between the BS against the black hole case, up to the maximum mass BS merger time, given by $t_{c,0.633}$ (i.e., $1-t/t_{c,0.633}\to0$).}}
    \label{fig:dephasing}
\end{figure*}

\begin{figure*}
    \includegraphics[width=1.5\columnwidth]{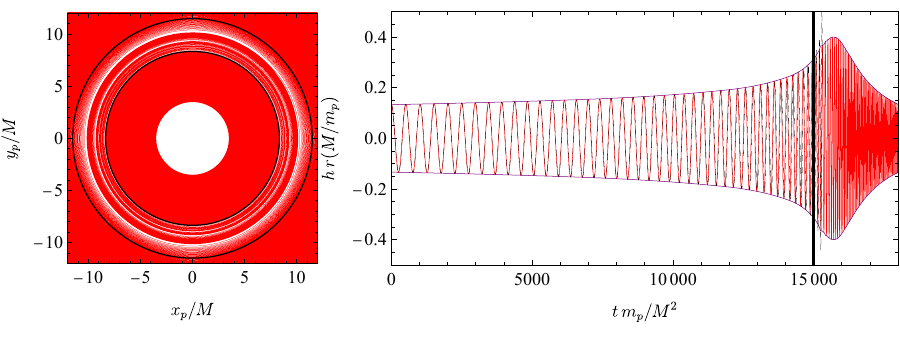}\\
    \includegraphics[width=1.5\columnwidth]{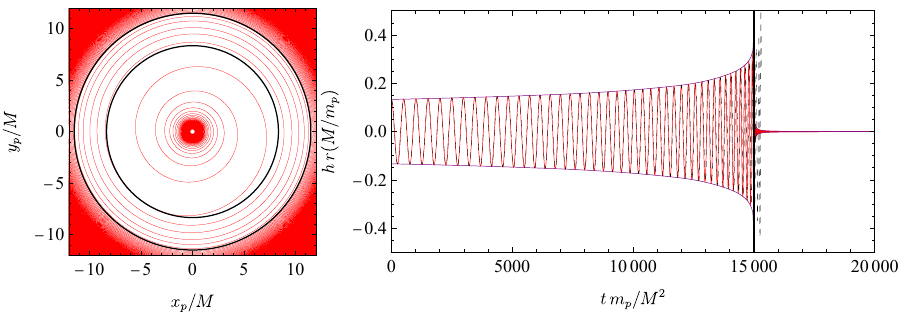}
    \caption{
Orbital motion (left) and gravitational waveform (right) for a particle inspiraling into a BS with no dipolar emission, for the case $M\mu = 0.62448$. We also show the BH case (dashed line) for comparison. In the left panels, the solid black circles indicate the thresholds for octupole and quadrupole scalar radiation activation.
Top panels: Scalar radiation has been artificially switched off, while keeping the resonant structure in the gravitational wave emission. This leads to resonant dephasing, visible as stripe-like features between the black lines representing the threshold for the octupolar and quadrupolar emission. Aside from these resonances, the orbital evolution proceeds smoothly, dominated by gravitational wave emission, and resembles the behavior predicted by the quadrupole approximation.
Bottom panels: As the particle crosses the radius where octupolar scalar radiation becomes active (outer black circle), the inspiral accelerates. Upon reaching the quadrupole threshold (inner black circle), the particle rapidly plunges toward the center. The resulting gravitational waveform exhibits a gradual decay with very small amplitude at late times, scaling approximately as $t^{-\alpha}$, where $\alpha\sim{\cal O}(10^{-2})$. }
    \label{fig:inspiral1}
\end{figure*}

\begin{figure*}
    \includegraphics[width=1.5\columnwidth]{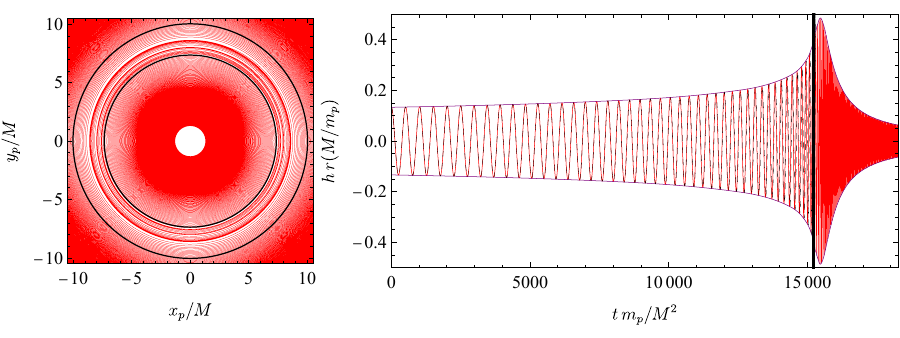}\\
    \includegraphics[width=1.5\columnwidth]{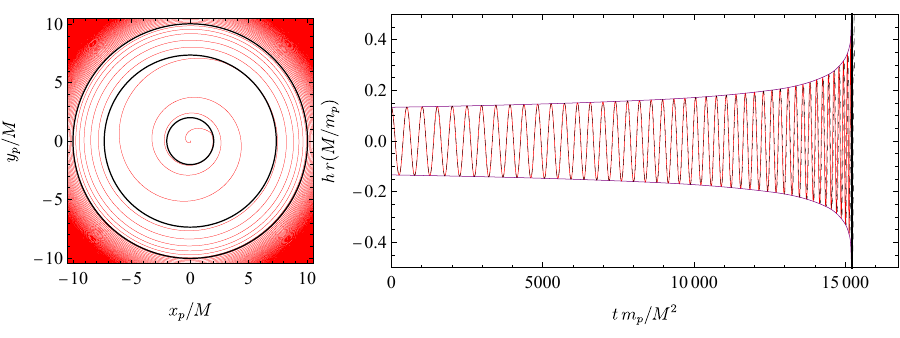}
    \caption{Same as Fig.~\ref{fig:inspiral1}, but with a central BS with mass $\mu M=0.633$, allowing for dipolar emission. We see that the overall picture is essentially the same as Fig.~\ref{fig:inspiral1}, with the exception that the dipolar radiation makes the inspiral to dive deep into the BS faster. This makes the overall waveform more similar that of a central BH.}
    \label{fig:inspiral2}
\end{figure*}

The similarity, seen in Fig.~\ref{fig:radius_evol}, between the inspiral into a BS and the Newtonian approximation for a binary system of point particles provokes us to read the BS as a  mimicker. In fact, for both BSs explored in our analysis, the radial evolution follows a behavior similar to
\begin{equation}
    r_p (t) \approx r_{\rm quad}=r_0\left(1-\frac{t}{t_c}\right)^{1/4},
\end{equation}
where $r_0 $ is the orbital radius at $t=0$, and $t_c$ is the coalescence time, given by
\begin{equation}
    t_{\rm c, quad}=\frac{5}{256}\frac{r_0^4}{M^2 m_p}.
    \label{eq:coal}
\end{equation}
In our computations, the expression above provides a good fit for the overall radial evolution up to the ``merger''. For the plots, however, we have adjusted the coalescence time so that $\approx 0.97 t_{\rm c, quad}$ in both cases. This shows that the discrepancy is indeed small, and can be partly attributed to relativistic corrections. Indeed, we have computed the inspiral of a point particle in the background of a Schwarzschild black hole and found the coalescence time to align more closely to the BS case than with the quadrupolar value given by Eq. \eqref{eq:coal}. 

Additionally, Fig.~\ref{fig:radius_evol} shows (as dotted blue curves) the evolution of the orbital radius when dynamical friction is neglected—i.e., excluding the scalar energy flux—so that the evolution is driven solely by gravitational radiation. This scenario shows a significantly slower inspiral, which highlights the considerable impact of dynamical friction on the orbital evolution. Although some dephasing still arises due to resonant configurations, it is much less pronounced than the effect coming from dynamical friction. Nonetheless, since BS-BS interactions are highly sensitive to the relative phase configuration, we find it instructive to include this comparison. In certain scenarios, phase differences can be tuned to substantially weaken the interaction between the stars, thereby greatly reducing the effects of dynamical friction \cite{Palenzuela:2006wp,Evstafyeva:2024qvp}.

Although the orbital radius evolves similarly across the different cases, the phase $\Psi(t)$ may exhibit noticeable discrepancies. This is expected, since the orbital frequency can vary significantly once the particle enters the relativistic regime and the motion extends to $r_p < 6M$. In Fig.~\ref{fig:dephasing}, we show the inspiral trajectory in the $(\Psi(t), r_p(t))$ plane, comparing the BS models with the BH case. For the configuration with $\mu M = 0.62448$, coalescence occurs earlier due to the larger effective radius of the BS, which enhances the efficiency of dynamical friction at larger separations. Nevertheless, even in this scenario, it is conceivable that a BH with a different mass could replicate the observed coalescence time, albeit not having the same phase evolution. 
% \RM{Isn't this worth exploring/easy to test?} \cm{Weakened a bit the statement} 
In contrast, the more compact configuration with $\mu M = 0.633$ yields an inspiral that closely resembles the black hole case. We argue that, in this regime, the signal can indeed mimic a BH chirp, which would make non-ultracompact BSs resemble black holes. 

Space-based LISA-like detectors are expected to observe EMRIs over many orbital cycles before the final plunge~\cite{Barack:2006pq,Babak:2017tow,Coogan:2021uqv}. Therefore, even if the resulting chirp-like signal reproduces all the qualitative features expected from a central BH, it is still worth computing the phase difference between the EMRI–BS and EMRI–BH cases. In the right panel of Fig.~\ref{fig:dephasing}, we show the accumulated phase difference during the inspiral evolution. We find that the presence of dipolar emission reduces the phase difference by an amount of order ${\cal O}(10M/m_p)$. Nonetheless, even when the signal exhibits a chirp-like behavior, its dependence on the mass ratio implies that LISA could still distinguish between a BS and a BH as the central object. Since LISA is expected to detect systems with $m_p/M \in (10^{-1},10^{-6})$, the possible dephasing for a BS–EMRI at coalescence can reach
\begin{equation}
\delta\Psi(t_c)\sim{\cal O}(10^1)\,\,{\rm to}\,\,{\cal O}(10^{6}),
\end{equation}
which lies well above the expected detectability threshold for LISA-type systems~\cite{Barack:2006pq,Babak:2017tow,Coogan:2021uqv}. Consequently, potential BS seeds could be identified well before the onset of the chirp-like phase. 

It is also worth pointing out that, in the absence of dipolar radiation, the phase evolution is related to the maximum of the angular frequency within the star, $\Omega_{\rm p,max}$. As the radiation gets feeble deep within the star, the emitted waves will have an approximately monochromatic behavior, with $\Psi(t)\approx 2\Omega_{\rm p,max} t+\delta$, where $\delta$ is some phase\footnote{This is a consequence of the evolution equation for the phase, which yields $\dot\Psi\approx 2\Omega_{\rm p,max}$ as $r\to 0$.}. This is a discriminator for stars with not enough compactness to activate the dipolar sector. We recall that long-lasting oscillations after the merger were observed in full numerical BS collisions in Ref. \cite{Croft:2022bxq}, but with a richer frequency spectrum. A similar feature happens when we neglect dynamical friction, with the inspiral driven solely by gravitational wave emission.

\begin{figure}
    \centering
    \includegraphics[width=1\linewidth]{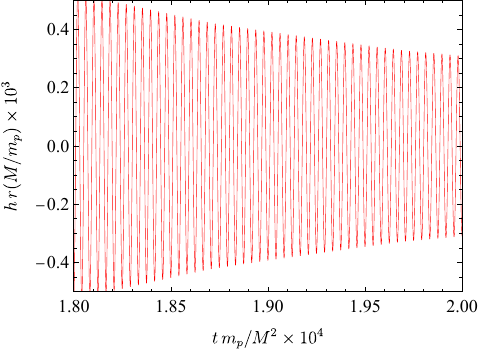}
    \caption{{Zoom-in of the late-time waveform shown in Fig.~\ref{fig:inspiral1}. Since the configuration considered here is very close to the onset of dipolar radiation, the waveform amplitude at late times is small.}}
    \label{fig:drag_zoom}
\end{figure}

It is instructive to examine the trajectories and waveforms directly. Once again, we stress that to better highlight the relevant effects, we extrapolate the perturbative results to the equal-mass regime. Although this extrapolation is not strictly valid, it provides a qualitatively accurate picture. Figures~\ref{fig:inspiral1} and~\ref{fig:inspiral2} present the orbital motion and gravitational wave amplitudes for the cases $\mu M = 0.62488$ and $\mu M = 0.633$, respectively. For comparison, we also include the black hole case in both figures, shown as dashed lines.
When dynamical friction is neglected (top panels), the inspiral proceeds gradually, driven solely by gravitational wave emission. The resonant behavior in the fluxes manifests as gaps in the orbital inspiral, corresponding to enhanced radiation at specific radial positions. As the particle enters the stellar interior, the inspiral continues at a slower pace, similar to the evolution observed in constant-density stars \cite{Macedo:2013qea}.
When dynamical friction is included, each active multipole channel (solid circles) contributes an additional drag force, accelerating the inspiral. In the less compact case (Fig. \ref{fig:inspiral1}), the absence of dipolar radiation allows for a smooth transition during the final stages of the inspiral, where the waveform at the end is approximately monochromatic, with (predominant) frequency $\sim 2\Omega_{\rm p,max}$, as stated before. {This behavior is shown more clearly in Fig.~\ref{fig:drag_zoom}, which displays a zoom-in of the late-time signal. We can see that the waveform remains regular and oscillatory rather than terminating abruptly}. In contrast, for the more compact configuration (Fig. \ref{fig:inspiral2}), dipolar radiation dominates and leads to a rapid plunge toward the center, abruptly ending the process.

\subsection{Backreaction on the stellar configuration}

\begin{figure}[h!]
    \centering
    \includegraphics[width=1\linewidth]{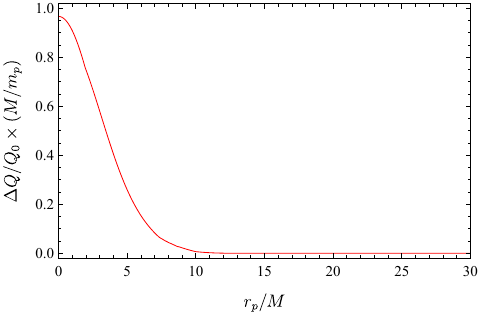}
    \caption{Total number of particles irradiated from the BS during the inspiral. We consider the maximum mass configuration with $M\mu=0.633$. Even in this extreme case, the total mass lost by the star is negligible in the EMRI as it scales with $m_p/M$.}
    \label{fig:fluxQ}
\end{figure}
 
The perturbative approach adopted here assumes that the central object, i.e., the BS, does not change considerably during the inspiral. Clearly, there must be at least some small evolution, as we have highlighted that the the perturber sheds matter, emitting it away to infinity. The validity of our assumption can be assessed by tracking the total number of scalar particles emitted during the inspiral, as given by Eq.~\eqref{eq:Q_ejected}. Figure~\ref{fig:fluxQ} depicts this quantity (normalized by the initial number of scalar particles) as a function of the orbital radius of the perturbing particle, for the marginally stable BS configuration (i.e., with $\mu M=0.633$)---the case in which the particle ejection is stronger. The total change in the number of particles is ${\cal O}(m_p/M)$, ensuring a negligible BS evolution through the inspiral. 

As briefly discussed above, besides shedding scalar field away [whose contribution is accounted for by the $\dot{E}_\infty^{\rm BS}$ term in the balance law \eqref{eq:flux}], the perturber also induces a quasi-static, spherically symmetric deformation of the BS. This deformation modifies the BS energy by $\delta E^{\rm BS}(r_p)\propto m_p M$. At the perturbative order considered in this work, the corresponding contribution~$ \dot{E}^{\rm BS}=\delta E^{\rm BS}{}'\dot{r}_{p}$ should therefore be included in the energy balance, adding to the $\dot{E}_{\rm orbit}$ in the left-hand side of Eq.~\eqref{eq:flux}. However, the Newtonian treatment in Appendix~\ref{app:backreaction} suggests that this effect can be completely absorbed in the mass of the particle, $m_p$.

\begin{figure}
    \centering
    \includegraphics[width=1\columnwidth]{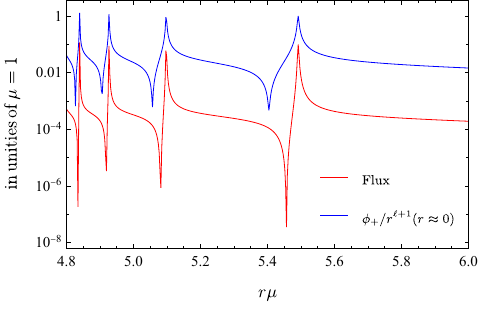}
    \caption{Scalar field perturbation computed near the center of the star as function of the orbital radius near the resonant configurations. We compare the amplitude of the scalar field near the center with the gravitational wave flux, highlighting that at the resonances there is an enhancement of the amplitude. }
    \label{fig:scal}
\end{figure}

The perturber can also deposit energy into the bosonic matter. While resonances of quasi-bound states are trapped and do not radiate scalar matter to infinity, they do induce changes in the scalar configurations that backreact into the particle motion. This was extensively explored in the case of gravitational atoms (see, e.g., \cite{Baumann:2018vus, Baumann:2019ztm, Berti:2019wnn, Tomaselli:2024bdd, Tomaselli:2024bdd}), and it is reasonable to expect that they play a role here as well. Considering particles moving through a soliton in a non-relativistic system, it has been shown that the trajectory can undergo a ``stone-skipping'' behavior, in which the orbital radius varies quasi-periodically due to the interaction between the particle and excitations of the field, in particular the dipolar mode~\cite{Wang:2021udl,Zhang:2026mse}. Therefore, the exchange of energy between the particle and the field profile can indeed be relevant. A way to implement in the relativistic setup this, would be to add the averaging stellar energy deposited by the perturber in the balance equation \eqref{eq:flux}. To illustrate how this feature is enhanced in the case of resonances, in Fig.~\ref{fig:scal} we show the scalar field perturbation amplitude near the center of the star, together with the flux to point the resonant peaks for the case $\mu M=0.633$. As we can see, the resonant configurations change the values of the field perturbations by orders of magnitude. A more detailed account on this effect will be left to future works.

\section{Discussion and conclusion}
\label{sec:conclusion}

In this work, we have investigated the inspiral of a compact object onto a supermassive BS within a fully relativistic perturbative framework. By modeling the system as an EMRI, we computed the gravitational and scalar fluxes responsible for the orbital evolution and the resulting waveform. Our analysis reveals that BSs can act as \emph{chirp mimickers}, producing signals that closely resemble those generated by black holes during the inspiral phase.

We have shown that the late-time dynamics of the inspiral depend critically on the compactness of the BS. For highly compact configurations, dipolar scalar radiation becomes active, enhancing the overall energy dissipation and accelerating the plunge of the secondary object. Less compact stars, on the other hand, only allow quadrupolar and higher-order scalar emissions, leading to a longer and smoother inspiral. Despite these differences, both cases exhibit chirp-like gravitational waveforms that could, at first glance, be confused with those from black hole binaries. However, we demonstrated that dephasing effects---arising from dynamical friction and resonant scalar excitations---are sufficiently large to be potentially resolvable by space-based detectors such as LISA.

We also derived semi-analytical prescriptions for the gravitational, scalar, and particle-number fluxes that accurately reproduce the numerical results even in the relativistic regime. These formulas provide simple and computationally efficient tools for exploring a wider range of BS models and inspiral parameters. Moreover, we confirmed that the backreaction on the star during the EMRI remains negligible, validating the perturbative and adiabatic assumptions employed throughout this work. It is noteworthy to point out that we disregard energy deposition into the star due to resonances, which are potentially relevant. 

Future research can extend the present analysis in several directions. Including self-interacting potentials and rotating BSs would provide a more realistic characterization of the scalar field configurations relevant for astrophysical applications. Incorporating higher-order self-force effects and eccentric or inclined orbits would improve the accuracy of waveform modeling and the assessment of detectability by LISA. Finally, exploring the interplay between resonant scalar modes and the evolution of the host star could shed light on possible observational imprints unique to bosonic compact objects.

\section*{Acknowledgments} 
CFBM and HCDL would like to thank Conselho Nacional de Desenvolvimento Científico e Tecnológico (CNPq), Coordenação de Aperfeiçoamento de Pessoal de Nível Superior (CA\-PES), Fundação Amazônia de Amparo a Estudos e Pesquisas (FAPESPA) and Fundação de Amparo à Pesquisa e ao Desenvolvimento Científico e Tecnológico do Maranhão (FAPEMA), from Brazil, for partial financial support. {HCDL also acknowledges the financial support of the Federal University of Maranhão (UFMA) through the ``Apoio a Jovens Pesquisadores'' grant (AGEUFMA 09/2025).}
R.M.~acknowledges financial support from CNPq and FAPERJ (Fundação Carlos Chagas Filho de Amparo à Pesquisa do Estado do Rio de Janeiro), Grant E-26/204.589/2024.
RV gratefully acknowledges the support of the Dutch Research Council
(NWO) through an Open Competition Domain Science-M grant, project number OCENW.M.21.375.
The Center of Gravity is a Center of Excellence funded by the Danish National Research Foundation under grant No. DNRF184.
We acknowledge support by VILLUM Foundation (grant no. VIL37766).
V.C.\ is a Villum Investigator.  
V.C. acknowledges financial support provided under the European Union’s H2020 ERC Advanced Grant “Black holes: gravitational engines of discovery” grant agreement no. Gravitas–101052587. 
Views and opinions expressed are however those of the author only and do not necessarily reflect those of the European Union or the European Research Council. Neither the European Union nor the granting authority can be held responsible for them.
This project has received funding from the European Union's Horizon 2020 research and innovation programme under the Marie Sklodowska-Curie grant agreement No 101007855 and No 101131233.
\appendix

\section{Boson star background and boundary conditions}\label{app:bosonstar}

\begin{figure*}
    \centering
    \includegraphics[width=\linewidth]{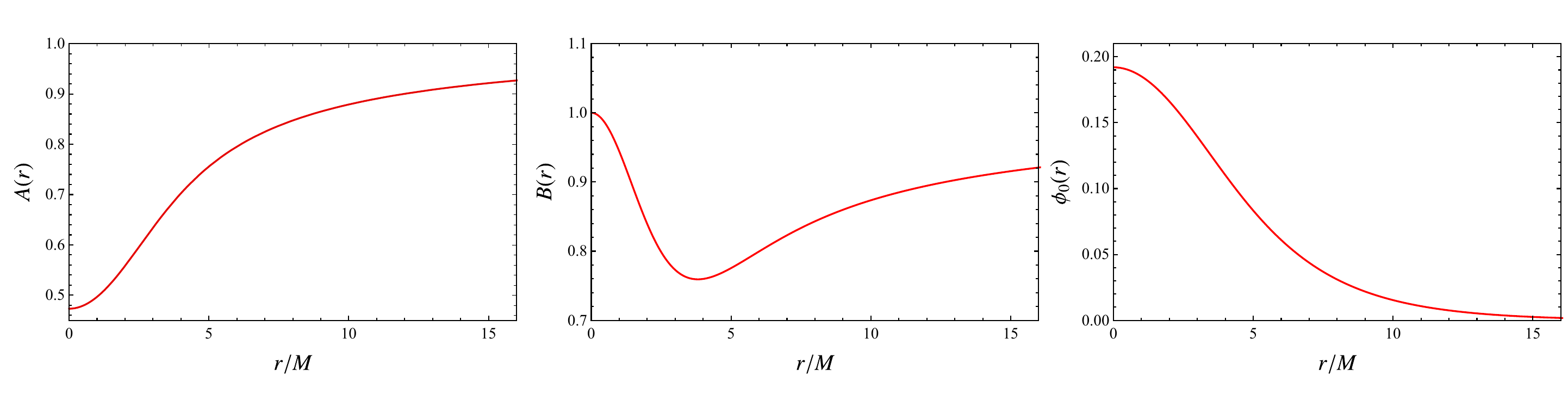}
    \caption{Numerical solutions for the metric components and the scalar field for the BS model. The numerical results correspond to the most compact configuration with total mass $M\mu=0.633$. }
    \label{fig:sols_num}
\end{figure*}

In this Appendix, we describe the background spacetime, the boundary conditions, and the numerical methods employed to obtain BS solutions. We consider spherically symmetric configurations in static equilibrium, with line element
\begin{align}
\label{line_el}ds^2=-A(r)\,dt^2+\frac{dr^2}{B(r)}+r^2\left(d\theta^2+\sin^2\theta d\varphi^2 \right),
\end{align}
where $A(r)$ and $B(r)$ are two real functions. The complex scalar field inherits the spherical symmetry of the background and can be written as
\begin{align}
\label{field_ans}\Phi(t,r)=\phi_0(r)\,e^{i\omega t},    
\end{align}
where $\phi_0(r)$ is a real function and $\omega>0$ is the scalar field frequency. Substituting the \textit{ansätze}~\eqref{line_el} and \eqref{field_ans} into the field equations~\eqref{eq:einstein} and \eqref{eq:scalar}, we obtain
\begin{align}
\label{fieldeq_1}&\frac{B'}{r} + \frac{(B - 1)}{r^2} = - 4 \pi \left( A^{-1} \omega^2 \phi_0^2 + B \phi'^{2}_0 + V \right),
\\
\label{fieldeq_2}& \frac{B A'}{r A} + \frac{(B - 1)}{r^2} = 4 \pi \left( A^{-1} \omega^2 \phi_0^2 + B \phi'^{2}_0 - V \right),
\\
\label{fieldeq_3}&
\phi_0'' + \left(\frac{2}{r} + \frac{A'}{2A} + \frac{B'}{2B}\right) \phi_0' + B^{-1}\left( A^{-1} \omega^2 - \frac{d V (|\Phi|^2)}{d|\Phi|^2} \right) \phi_0 = 0,
\end{align}
where primes denote derivatives with respect to the radial coordinate $r$. Equations~\eqref{fieldeq_1}-\eqref{fieldeq_3} form a system of coupled ordinary differential equations (ODEs) for the metric functions $A(r)$ and $B(r)$ and the radial part of the scalar field, $\phi_0(r)$. 

For a given potential $V(|\Phi|^2)$, asymptotically flat and regular BS configurations can be obtained by imposing the following boundary conditions: 
\begin{align}
\label{bc1}&B(0)=1, \qquad \lim_{r\to\infty}A(r)=1\\
\label{bc2}&\phi'_0(0)=0, \qquad \lim_{r\to\infty}\phi_0(r)=0.
\end{align}
The system of coupled ODEs~\eqref{fieldeq_1}-\eqref{fieldeq_3}, subjected to conditions~\eqref{bc1} and \eqref{bc2}, constitutes a boundary value problem (BVP). We numerically solve this BVP using a standard shooting method, i.e., reducing it to an initial-value problem (IVP). The function $A(r)$ can be rescaled after the integration such that it goes asymptotically to Schwarzschild as $r\to\infty$.

For concreteness, we adopt in this paper the mini-boson star model described by the effective potential:
\begin{align}
V(|\Phi|^2)=\mu^2|\Phi|^2,
\end{align}
which is the simplest model admitting solitonic solutions in General Relativity. 
In this case, the field equations \eqref{fieldeq_1}-\eqref{fieldeq_3} can be further simplified by the following rescaling:  
\begin{align}
r\rightarrow \frac{\tilde{r}}{\mu}, \quad \omega \rightarrow \tilde{\omega}\mu, \quad \phi_0(r) \rightarrow \frac{\tilde{\phi}_0(r)}{\sqrt{4\pi}}.
\end{align}

In Fig.~\ref{fig:sols_num}, we show an example of a numerical solution corresponding to the most compact configuration in the mini-boson star model, with total mass $\mu M=0.633$. We present the metric components $A(r)$ (left panel) and $B(r)$ (middle panel), as well as the scalar field profile $\phi_0 (r)$ (right panel), as a function of the rescaled radial coordinate.

It is clear that BSs lack a well-defined surface, at which pressure vanishes. Instead, the scalar field decays exponentially at large distances from the center. In practice, we define an effective radius for the BS, as the radial coordinate $R$ that encloses $99\%$ of its total mass. If we define the mass function $m(r) = (r/2) (1 - B)$, then the total mass of the star is given by $M = \lim_{r\to\infty} m(r)$ and the effective radius is defined through
\begin{equation}
    m(R)=0.99 M.
\end{equation}
While the definition of the radius is artificial, it traces a boundary beyond which effects of the scalar field become less important. We stress that, in this paper, we do not assume the spacetime to be Schwarzschild immediately outside the (effective) radius $R$ of the star. Instead, we run the numerical integration up to a region where $\phi(r)\sim 10^{-6}\phi_c$, with $\phi_c$ being the central value of the scalar field. Only at that point do we glue the BS spacetime to the Schwarzschild one.

%%%%%%%%%%%%%%%%%%%%%%%%%%%%%%%%%%%%%%%%%%%%%
\section{Nonradial perturbations of boson stars}\label{app:perturbations}
%%%%%%%%%%%%%%%%%%%%%%%%%%%%%%%%%%%%%%%%%%%%
Here, for completeness, we outline the main procedure to find the equations describing the perturbations due to an orbiting particle around a BS. Details can also be found in other references, such as \cite{1967ApJ...149..591T,Macedo:2013jja,Duque:2023seg,Duque:2024fot}. We focus on the polar sector, as this is dominant for circular orbits and is the sector that contributes to dynamical friction.

To expand the polar sector, we can use the Regge-Wheeler gauge to express the metric perturbation as
\begin{align}
   ds^2_{\rm polar}&=\int d\sigma e^{-i\sigma t}Y_{\ell m} (\theta,\varphi) {\bigg[}A H_0dt^2 \nonumber\\
   &+2i\sigma H_1 dtdr+\frac{H_2}{B}dr^2+ Kd\Omega^2 {\bigg]},
\end{align}
where $(H_0,H_1,H_2,K)$ are functions of the radial coordinate only, and $Y_{\ell m}(\theta,\varphi)$ are the spherical harmonics. In the expressions for the perturbations, we are implicitly assuming a sum over the harmonics index $\ell\geq|m|$. The scalar field can be expanded in terms of harmonics. A useful way to express it is~\cite{Macedo:2013jja,Yoshida:1994xi} 
\begin{align}
    \delta\Phi&=\int d\sigma\frac{\phi_+}{r}Y_{\ell m} (\theta,\varphi)e^{-i(\sigma +\omega)t},\\
    \delta\Phi^*&=\int d\sigma\frac{\phi_-}{r}Y_{\ell m} (\theta,\varphi)e^{-i(\sigma -\omega)t},
\end{align}
where, once again, $\phi_\pm$ are function of the radial coordinate, only. Finally, we can decompose the particle's stress-energy tensor in terms of tensor harmonics. In the time domain, the stress-energy tensor can be written as
\begin{equation}\label{eq:particlestress}
    T_{{\rm p,}\mu\nu}=\sqrt{\frac{A}{B}}m_p\frac{\dot{x}_{p,\mu}\dot{x}_{p,\nu}}{r_p(t)^2\dot{x}_{p}^t}\delta(r-r_p(t))\delta(\cos\theta)\delta(\varphi-\varphi_p).
\end{equation}
We can find the Fourier coefficients of the above as well as a spherical harmonic decomposition in terms of spherical tensor harmonics~\cite{Thorne:1980ru}. We  can write
\begin{align}
    {\bf T}_{\rm p} &=
A_{\ell m}^{(0)} {\bf a}_{\ell m}^{(0)}
+ A_{\ell m}^{(1)} {\bf a}_{\ell m}^{(1)}
+ A_{\ell m} {\bf a}_{\ell m}
+ B_{\ell m}^{(0)} {\bf b}_{\ell m}^{(0)}\nonumber\\
&+ B_{\ell m} {\bf b}_{\ell m}
+ Q_{\ell m}^{(0)} {\bf c}_{\ell m}^{(0)}
+ Q_{\ell m} {\bf c}_{\ell m}\nonumber\\
&+ D_{\ell m} {\bf d}_{\ell m}
+ G_{\ell m}^{(s)} {\bf g}_{\ell m}
+ F_{\ell m} {\bf f}_{\ell m},\label{eq:tensorharmonic}
\end{align}
where the $({\bf a},{\bf b},{\bf c},{\bf d},{\bf f},{\bf g})$ terms are the (matrix representation of) tensor harmonics, which can be explicitly seen in Ref.~\cite{Zerilli:1970se,Sago:2002fe}. By equating Eqs.~\eqref{eq:particlestress} and~\eqref{eq:tensorharmonic}, and using the orthogonality conditions, we obtain the explicit form of the coefficients in the expansion~\eqref{eq:tensorharmonic} in terms of the particle position. The full expression is shown Table I of Ref.~\cite{Sago:2002fe}. %We notice here that, since we are using only the polar sector, we 

Once we have the metric, scalar field and particle quantities in terms of tensor spherical harmonics we can use the perturbed Einstein's equation to compute the basic equations to be integrated.

\section{Comparison between the Newtonian fit and the numerical computations}\label{app:approximation}

\begin{figure}
    \centering
    \includegraphics[width=0.8\linewidth]{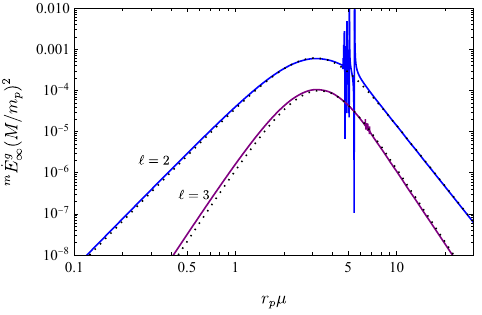}
    \caption{Comparison between the numerically computed gravitational flux [Eq.~\eqref{eq:flux_Eg}] (solid lines) and the modified Newtonian approximation in Eq.~\eqref{eq:mod_flux} (dotted lines). Here we consider the maximum mass configuration, but we notice that the agreement is even better in low-compactness scenarios.}
    \label{fig:Num_vs_an}
\end{figure}

\begin{figure}
    \centering
    \includegraphics[width=0.8\linewidth]{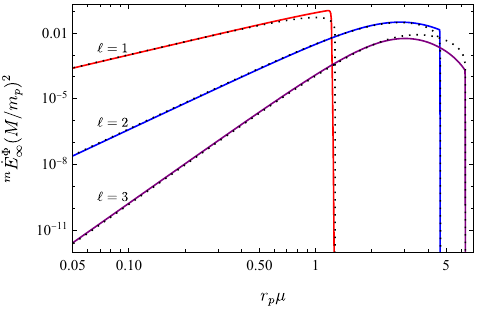}
    \caption{Comparison between the numerically computed scalar flux, Eq.~\eqref{eq:flux_EPhi} (solid lines), and the modified Newtonian approximation, Eq.~\eqref{eq:scalar_flux} (dotted lines), considering the marginally stable BS configuration, with $\mu M=0.633$. For the dipolar case, $\ell=m=1$, even though the approximation is not expected to hold, introducing an additional factor of $\sim 2$ in the formula brings the flux into agreement with the numerical result. For small values of $r_p$ there is good agreement between the full and semi-analytic results.}
    \label{fig:num_vs_an2}
\end{figure}

\begin{figure}
    \centering
    \includegraphics[width=0.8\linewidth]{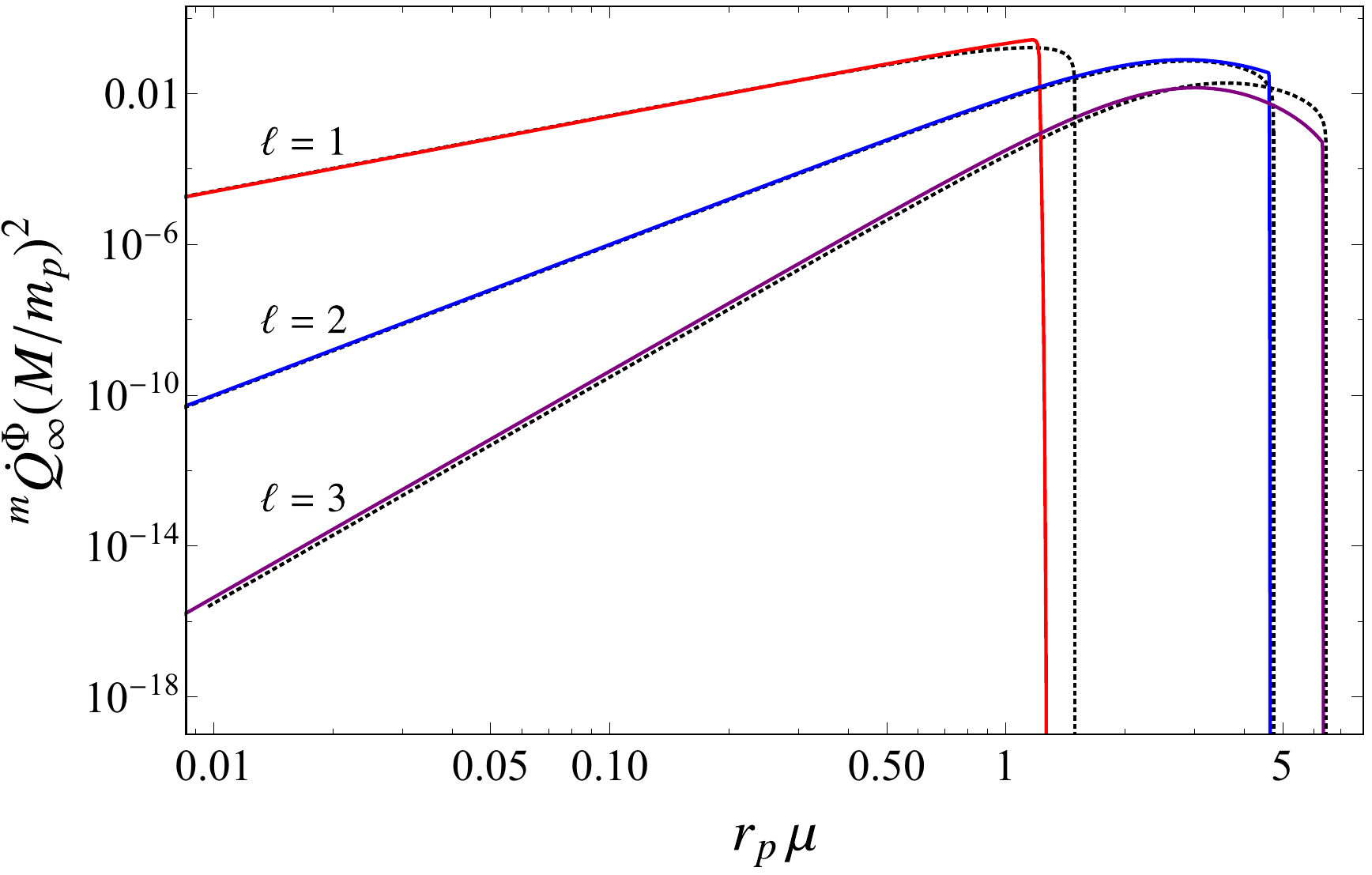}
    \caption{Comparison between the numerically computed particle number flux, given in Eq.~\eqref{eq:flux_QPhi} (solid lines), and the semi-analytic approximation (dotted lines), considering a BS configuration with $\mu\,M=0.633$. For small values of $r_p$ the semi-analytic and numerical results agree very well. For the dipolar case $l=m=1$ an additional factor of $\sim 2$ was introduced in the adjustment.}
    \label{fig:number_flux}
\end{figure}

In this section, we show a numerical comparison between the fluxes computed using the full system of equations and the ones obtained using the modified Newtonian approximation. The results demonstrate that the approximation can capture the overall behavior of the inspiral fluxes.

In Fig. \ref{fig:Num_vs_an} we show a comparison between the full numerical computation and the semi-analytical fit for the gravitational flux. We see that aside from the resonant peaks, the fit approximates well the gravitational wave flux even in the strong field regime. We focus on a high-compactness configuration, but the fit is also good (even better) for lower compactnesses as $g_{tt}\approx 1$ in those cases. 

We note that the resonant peaks could, in principle, be incorporated into the modified Newtonian formula, provided the quasinormal modes of the stellar configurations are known. This type of fit has previously been explored for EMRIs around neutron stars, motivated by the analogy with a simple forced harmonic oscillator~\cite{Kojima:1987tk,Pons:2001xs}. We do not explore this possibility here.

In Fig.~\ref{fig:num_vs_an2} we compare the scalar fluxes and in Fig.~\ref{fig:number_flux} we compare the particle fluxes, for $\ell=1,2, 3$, with the modified Newtonian approach. We notice that in both cases the Newtonian expression fits well the numerical results in the region $r_p \sim  \mathcal{O}(R)$, although in the dipolar case ($l=m=1$) one needs to introduce a factor of $2$ into the fit. This is noteworthy as the Newtonian approximation for the scalar field perturbation was deduced for the cases with $\ell\geq 2$.

\section{Validity of the quasi-circular motion}\label{app:qc}

Due to the rapid evolution of the inspiral driven by scalar radiation, it is natural to question whether the quasi-circular adiabatic approximation remains valid. A key quantity to assess this is
\begin{equation}
\delta_p=\frac{\dot{\Omega}_p}{\Omega_p^2}=\frac{\Omega_p'}{\Omega_p^2}\dot{r}_p=-\frac{\Omega_p'}{\Omega_p^2}\frac{\mathcal{F}}{E_{\rm orbit}'},
\end{equation}
where Eq.~\eqref{eq:drorbit} has been used to eliminate $\dot{r}_p$. For quasi-circular motion, we require $\delta_p \ll 1$. This condition naturally depends on the mass ratio $m_p/M$ and is always satisfied as long as $m_p/M \ll 1$.

\begin{figure}
    \centering
    \includegraphics[width=0.8\linewidth]{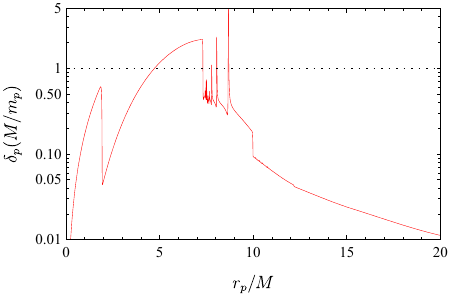}
    \caption{Quasi-circular discriminator for the adiabatic evolution. As long as $\delta_p\ll 1$, the quasi-circular approximation should be valid.}
    \label{fig:adiabatic}
\end{figure}

In Fig.~\ref{fig:adiabatic}, we show the behavior of $\delta_p$ as a function of the particle’s orbital radius. We find that, for $m_p/M=1$, $\delta_p > 1$ only occurs near the region where quadrupolar scalar emission is activated. Deep inside the star, even at locations where dipole radiation is active, we obtain $\delta_p < 1$ for equal-mass binaries. If we restrict ourselves to fiducial values of $\delta_p < 0.1$ outside of the resonances, this corresponds to $m_p/M <0.04$, which we take as a reference for the validity of the quasi-circular approximation.

\section{Backreaction on the BS configuration in the Newtonian limit}
\label{app:backreaction}

Here, we develop some of the points discussed in the text about the backreaction on the stellar configuration, focusing on the Newtonian limit. 

It can be easily shown that the equality $E^{\rm BS}=\omega Q$ holds for any stationary configuration, including in the relativistic regime (e.g.~\cite{Brito:2023pyl}). A monopole static perturbation from an orbiting body near a BS makes the eigenvalue $\omega(Q)$ dependent on $m_p/M$ and $r_p \mu$, i.e.,
\begin{equation}
   \omega\approx\omega_0(Q)+\frac{m_p}{M}\, \Delta(Q,r_p \mu)\,,
\end{equation}
where $\omega_0$ is the background BS eigenvalue, and $\Delta$ is a numerical function obtained by solving the perturbed BS at linear order in $m_p/M$. We now derive these functions in the Newtonian limit.

First, we consider the background NBS. The Schrödinger-Poisson equation for the NBS has a well-known scaling symmetry (e.g.~\cite{Annulli:2020lyc}),
\begin{equation}
   \Big(\psi_0(r M \mu^2),U_0(r M \mu^2),\omega_0-\mu\Big)\propto Q^2\,, \quad M\approx \mu Q\,.
\end{equation}
This implies $d\omega_0=2 (\omega_0-\mu)\,d Q/Q$. Therefore, in the absence of a perturber: $d E^{\rm NBS}=[\omega_0+2(\omega_0-\mu)]\,d Q$.

We turn now to the NBS quasi-static monopole deformation; these perturbations satisfy the linearized equations~\cite{Annulli:2020lyc}
\begin{gather}
 \frac{1}{r^2} \partial_r\left[ r^2 \partial_r\delta\psi\right] = 2\mu^2\left\{\left[U_0-(\omega_0-\mu) \right] \delta\psi+\psi_0\, \delta U\right\}\,, \label{eq:SP_lin_sch}\\
 \frac{1}{r^2} \partial_r\left[ r^2 \partial_r\delta U\right] = 4\pi \left[ 2  \mu^2\psi_0\, \delta\psi+\frac{m_p}{r^2}\delta(r-r_p)\right] \,, \label{eq:SP_lin_pois}
\end{gather}
where $\psi_0$ and $U_0$ are the field and gravitational potential of the background NBS. This system must be solved imposing regularity at the origin and at spatial infinity. The charge contained in such perturbations ($\delta Q_p$) is generally non-vanishing.

As noted in~\cite{Annulli:2020lyc}, a (time-dependent) monopole homogeneous solution
\begin{equation}
  \delta \psi_{\varepsilon} = \varepsilon\, \psi_0 \left[1 - i (\omega_0-\mu) t\right]\,, \qquad \delta U_{\varepsilon} = \varepsilon\, U_0 \,,
\end{equation}
with $\epsilon\ll1$, can be added to the $(\delta \psi,\delta U)$-solution. This describes a (infinitesimal) change in the background NBS configuration, with charge and frequency altered by
\begin{equation}
  \delta Q_\epsilon= \frac{\epsilon}{2} Q\,, \qquad \delta \omega_\epsilon=\epsilon\, (\omega_0-\mu)\,.
\end{equation}

Fixing the parameter $\epsilon\equiv-2\delta Q_p/Q$ by requiring that the total charge contained in the perturbation $(\delta \psi+\delta \psi_\epsilon,\delta U+\delta U_\epsilon)$ vanishes, we obtain
\begin{equation}
    \Delta=-2 (\omega_0-\mu) \frac{\delta Q_p}{(m_p/M)Q}\,,
\end{equation}
with $\delta Q_p/(m_p/M)$ only a function of $x\equiv r_p M \mu^2$.
At linear order in $m_p/M$, the NBS quasi-static monopole deformation contributes to $dE^{\rm NBS}$ through
\begin{equation}
    dE^{\rm NBS}_p=Q\,d\left[\frac{m_p}{M}\, \Delta\right]\approx0.32\,(\mu M)^4\frac{m_p}{M} \left[\frac{\delta Q_p}{m_p/M}\right]'d r_p\,,
\end{equation}
where the prime denotes a derivative with respect to $x$, and we used $(\mu-\omega_0)/(M^2\mu^3)\approx 0.16$ (e.g.~\cite{Annulli:2020lyc}).
Remarkably, our numerics indicate that $\delta Q_p/(m_p/M)\approx(\mu M)^{-2}(U_0+\frac{x}{2}U_0')'$ to an error $\lesssim 1\%$.
Thus, from the relation
\begin{equation}
    d E_{\rm orbit}=(\mu M)^4 \frac{m_p}{M}  \left[\frac{U_0 + \tfrac{x}{2} U_0' }{(\mu M)^2}\right]' d r_p\,,
\end{equation}
we find $dE_p^{\rm NBS}\approx0.32\, dE_{\rm orbit}$.

The final energy balance equation accounting for all the above effects in the Newtonian regime is
\begin{equation}
    \dot{E}_{\rm orbit}\approx\tfrac{1}{1.32}\left[\dot E^g_\infty + [m\Omega_p+2(\mu-\omega_0)]\dot{Q}_\infty \right]\,.
\end{equation}

\bibliography{biblio}

\end{document}